\input harvmac
\def\O{{\cal O}}
\def\l{\lambda}
\def\e{\epsilon}
\def\ep{\epsilon^\prime}
\def\ra{\rightarrow}
\def\gsim{{~\raise.15em\hbox{$>$}\kern-.85em
          \lower.35em\hbox{$\sim$}~}}
\def\lsim{{~\raise.15em\hbox{$<$}\kern-.85em
          \lower.35em\hbox{$\sim$}~}}
\def\Im{{\rm Im}}
\def\Re{{\rm Re}}

\Title{hep-ph/9607415, WIS-96/31/Jul-PH}
{\vbox{\centerline{On Supersymmetric CP Violation}}}
\bigskip
\centerline{Yosef Nir}
\smallskip
\centerline{\it Department of Particle Physics,
 Weizmann Institute of Science, Rehovot 76100, Israel}
\bigskip
\bigskip
\baselineskip 18pt

\noindent
We discuss the discovery potential for New Physics of various
measurements of CP violation. If nature is supersymmetric, then
the flavor problem is even more mysterious than in the standard model.
We show how we can learn about the mechanism that solves the
supersymmetric flavor problem from measurements of mixing and CP
violation in $K$, $D$ and $B$ decays.
\bigskip
\bigskip

\centerline{\it Talk presented at the Workshop on $K$ Physics}
\centerline{\it Orsay, France, May 30 -- June 4, 1996}

\Date{7/96}

\newsec{Introduction}

One of the most intriguing aspects of high energy physics is
CP violation. On the experimental side, it is one of the
least tested aspects of the Standard Model. There is only
one (complex) CP violating parameter that is unambiguously measured
\ref\first{J.H. Christenson, J.W. Cronin, V.L. Fitch and R. Turlay,
Phys. Rev. Lett. 13 (1964) 138.},
that is the $\e$ parameter in the neutral $K$ system.
A genuine testing of the Kobayashi-Maskawa picture of CP violation
\ref\KoMa{M. Kobayashi and T. Maskawa, Prog. Theo. Phys. 49 (1973) 652.}\
in the Standard Model awaits the building of $B$ factories that
would provide a second, independent measurement of CP violation
\ref\BCP{A.B. Carter and A.I. Sanda, Phys. Rev. Lett. 45 (1980) 952;
 Phys. Rev. D23 (1981) 1567.}.
On the theoretical side, the Standard Model picture of CP violation
has two major difficulties. First, CP violation is necessary for
baryogenesis, but the Standard Model CP violating processes are
much too weak to produce the observed asymmetry of the Universe.
Simple extensions of the Standard Model do provide large enough sources
of CP violation that can be consistent with the observed asymmetry
\ref\CKN{For a review, see A.G. Cohen, D.B. Kaplan and
A.E. Nelson,  Ann. Rev. Nucl. Part. Sci. 43 (1993) 27.}.
Second, an extreme fine-tuning is needed in the CP violating part
of the QCD Lagrangian in order that its contribution to the electric
dipole moment of the neutron does not exceed the experimental upper
bound. This suggests that an extension of the Standard Model, such
as a Peccei-Quinn symmetry
\ref\PeQu{R.D. Peccei and H.R. Quinn, Phys. Rev. Lett. 38 (1977) 1440;
Phys. Rev. D16 (1977) 1791.}\
or a horizontal symmetry that guarantees $m_u=0$
\ref\BNS{For a review, see T. Banks, Y. Nir and N. Seiberg,
 hep-ph/9403203.},
is required.

The implications of baryogenesis for CP violation are particularly
interesting. GUT baryogenesis
\ref\KoTu{For a review, see E.W. Kolb and M.S. Turner,
{\it The Early Universe}  (Addison-Wesley 1990).},
while providing very plausible mechanisms for Sakharov's requirements
\ref\Sakh{A.D. Sakharov, ZhETF Pis. Red. 5 (1967) 32;
 JETP Lett. 5 (1967) 24.}\
($B$ nonconserving interactions, violation of both C and CP, and
departure from thermal equilibrium), runs into serious difficulties.
In particular, any baryon asymmetry produced prior to inflation is washed
out by inflation. For GUT scale baryogenesis to occur after inflation
requires a high reheat temperature $T_{\rm rh}$. Constraints from
structure formation, $T_{\rm rh}\lsim10^{12}\ GeV\ll m_{\rm GUT}$ and
(within supergravity models) from Nucleosynthesis constraints,
$T_{\rm rh}\lsim10^{10}\ GeV\ (m_{\rm grav}/100\ GeV)$,
make this unlikely.
Moreover, electroweak processes at $T=\O(TeV)$ might completely wash out
an earlier generated baryon asymmetry with initially vanishing $B-L$.
These problems suggest that the processes that are responsible
to the observed baryon asymmetry have taken place at temperatures
of the order of the electroweak scale \CKN.

Remarkably, the Standard Model itself has the potential of dynamically
generating baryon asymmetry
\ref\FaSh{G.R. Farrar and M.E. Shaposhnikov, Phys. Rev. D50 (1994) 774.}.
However, departure from thermal equilibrium
can only occur at the electroweak epoch if there is a sufficiently strong
first order phase transition. This requires a light SM Higgs, below
the experimental bound, or an extension of the Higgs sector.
More important to our discussion is the fact that CP violation
in the Standard Model is far too small
\nref\Gave{M.B. Gavela {\it et al.}, Nucl. Phys. B430 (1994) 382.}%
\nref\HuSa{P. Huet and E. Sather, Phys. Rev. D51 (1995) 379.}%
\refs{\Gave-\HuSa}. It allows
at best $n_B/s\simeq10^{-20}$, and perhaps a lot less.
Simple extensions of the Standard Model, such as the Minimal
Supersymmetric Standard Model or Two Higgs Doublet Models,
have extended Higgs sectors that do allow a first order
phase transition, and new sources of CP violation that
could be consistent with the observed $n_B/s$, but only if
the new phases are not too small.

The conclusion then is that it is not unlikely that there
exist {\it large, new sources of CP violation at the
electroweak scale}. This makes the experimental search for
CP violation in all its possible low energy manifestations
a very exciting direction of research.
We note, however, that CP violating phases that can account
for the baryon asymmetry are most likely to be probed in measurements
of electric dipole moments. It is very difficult to induce
large enough baryogenesis with flavor dependent phases of the type that
may affect CP asymmetries in $B^0$ decays.

In this work, we focus on supersymmetry as an example of
New Physics which potentially affects CP violation. We will
discuss in detail CP violation in neutral meson mixing.
We will not discuss the implications of supersymmetry
on electric dipole moments. We would like to mention, however,
that supersymmetric theories have at least two new flavor-diagonal
CP violating phases
\nref\DGH{M. Dugan, B. Grinstein and L. Hall, Nucl. Phys. B255 (1985)
413.}%
\nref\DiTh{S. Dimopoulos and S. Thomas, Nucl. Phys. B465 (1996) 23.}%
\refs{\DGH,\DiTh}. While these phases could
generate the observed baryon asymmetry
\ref\HuNe{For a recent study, see P. Huet and A.E. Nelson,
 Phys. Rev. D53 (1996) 4578.},
they also typically give an electric dipole moment of the neutron that is
two orders of magnitude above the experimental bound. Most supersymmetric
models simply fine tune the new phases to zero (though models with
naturally small phases have been constructed
\nref\BaBa{K.S. Babu and S.M. Barr, Phys. Rev. Lett. 72 (1994) 2156.}%
\nref\MaRA{M. Masip and A. Rasin, Nucl. Phys. B460 (1996) 449.}%
\nref\BDH{R. Barbieri, G. Dvali and L.J. Hall,
 Phys. Lett. B377 (1996) 76.}%
\nref\NiRa{Y. Nir and R. Rattazzi, hep-ph/9603233.}%
\nref\MoRa{R.M. Mohapatra and A. Rasin, hep-ph/9604445.}%
\nref\BuRr{K.S. Babu and S.M. Barr, hep-ph/9606384.}%
\nref\DNS{M. Dine, Y. Nir and Y. Shirman, hep-ph/9607397.}%
\refs{\BaBa-\DNS,\DiTh}). If supersymmetry exists in Nature, and
if the supersymmetric phases are indeed responsible for baryogenesis,
the phases cannot be much below the bound. This means that the on-going
search for $d_N$ may well yield a signal. Alternatively, improved upper
bounds on $d_N$ become
more and more of a serious problem to the supersymmetric framework.

\nref\NIR{For a review, see
Y. Nir, in {\it Proc. of the 20th Annual Slac Summer
Institute on Particle Physics:  The Third Family and the Physics of
Flavor}, Stanford, CA (1992), p. 81.}%
\newsec{CP Violation in Neutral Meson Systems \NIR}

We are mainly interested in pairs of decay processes
that are related by a CP transformation. If $B$ and $\bar B$
are CP conjugate mesons and $f$ and $\bar f$ are CP conjugate
states, then we denote by $A$ and $\bar A$ the two CP conjugate
decay amplitudes:
\eqn\AbarA{A=\vev{f|H|B},\ \ \ \bar A=\vev{\bar f|H|\bar B}.}
We define $p$ and $q$ ($|p|^2+|q|^2=1$) as the components of
the two neutral interaction eigenstates $B^0$ and $\bar B^0$ in the
mass eigenstates $B_1$ and $B_2$:
\eqn\pandq{|B_1\rangle=p|B^0\rangle+q|\bar B^0\rangle,\ \ \
|B_2\rangle=p|B^0\rangle-q|\bar B^0\rangle.}
We define a quantity $\lambda$,
\eqn\deflam{\lambda={q\over p}{\bar A\over A}.}
The three quantities $|\bar A/A|$, $|q/p|$ and -- for final CP
eigenstates -- $\lambda$ are independent of phase conventions and
correspond to three distinct types of CP violation.

(i) CP violation in decay:
\eqn\indecay{|\bar A/A|\neq1.}
This is a result of interference between various decay amplitudes
that lead to the same final state. It can be observed in charged
meson decays. The processes that are likely to have
non-negligible effects are decays with suppressed tree contributions,
e.g. $B\ra\rho K$, decays with no tree contributions, e.g.
$B\ra\phi K$ and $B\ra KK$, and radiative decays.
A theoretical calculation of this type of CP violation,
\eqn\thedec{\left|{\bar A\over A}\right|=\left|{
\sum_i A_i e^{i\delta_i}e^{-i\phi_i}\over
\sum_i A_i e^{i\delta_i}e^{+i\phi_i}}\right|,}
requires knowledge of strong phase shifts $\delta_i$ and absolute values
of amplitudes $A_i$ to extract the weak, CP violating phases $\phi_i$.
Consequently, it involves large hadronic uncertainties.

(ii) CP violation in mixing:
\eqn\inmix{|q/p|\neq1.}
This is a result of the mass eigenstates being
non-CP eigenstates. It can be observed in semileptonic neutral
meson decays. A theoretical calculation of this type of CP violation,
\eqn\themix{\left|{q\over p}\right|^2=\left|{
M_{12}^*-{i\over2}\Gamma_{12}^*\over
M_{12}-{i\over2}\Gamma_{12}}\right|,}
requires knowledge of $B_K$ in the $K$ system or $\Gamma_{12}$ in the $B$
system. Consequently, it involves large hadronic uncertainties.

(iii) CP violation in the interference of mixing and decay:
\eqn\ininter{\lambda\neq1.}
In particular, we mean here $|\lambda|=1$ and $\Im\lambda\neq0$.
This is a result of interference between the direct decay into a final
state and the first-mix-then-decay path to the same final state. It can
be observed in decays of neutral mesons into final CP eigenstates.
A theoretical calculation of this type of CP violation could be
theoretically very clean, provided that two conditions are met:
\item{a.} $A$ is dominated by a single weak phase, so that
CP violation in decay has no effect.
\item{b.} $\Im\lambda\gg10^{-3}$, so that the effect of CP
violation in mixing is negligible.

The $K\ra\pi^+\pi^-$ decays satisfy the first condition, but
$\Im\lambda(K\ra\pi\pi)\sim10^{-3}$, which is the reason why
we do not have a very clean determination of the Kobayashi-Maskawa
phase from the $K$ system. On the other hand, both conditions are
satisfied in various $B$ decays, e.g. $B\ra\psi K_S$ and (with
isospin analysis) $B\ra\pi\pi$. This is why $B$ factories would enable us
to determine $\sin2\alpha$ and $\sin2\beta$ very cleanly.

We conclude that CP asymmetries in neutral $B$ decays are a unique
tool for discovering New Physics: due to their theoretical cleanliness,
they are sensitive to New Physics even if it gives a contribution that
is comparable to the Standard Model one. Other CP violating observables
in meson decays (and, similarly, the electric dipole moment of the
neutron) can clearly signal New Physics only if the new contribution is
much larger than that of the Standard Model.

\newsec{The $K$ System}

The smallness of Flavor Changing Neutral Current (FCNC) processes
(particularly $\Delta m_K$) and of CP violation (particularly
$\e$) in the $K$ system provides severe tests and puts stringent
constraints on extensions of the Standard Model. In this section we
discuss the impact of $K$ physics on supersymmetric models building.
But first, we explain which types of CP violation contribute to
$\e$ and to $\ep$.

\subsec{The $\e$ and $\ep$ Parameters}

The two CP violating quantities measured in the neutral $K$ system are
\eqn\defeta{
\eta_{00}={\vev{\pi^0\pi^0|H|K_L}\over\vev{\pi^0\pi^0|H|K_S}},\ \ \
\eta_{+-}={\vev{\pi^+\pi^-|H|K_L}\over\vev{\pi^+\pi^-|H|K_S}}.}
We define
\eqn\defApp{\eqalign{
A_{00}=\vev{\pi^0\pi^0|H|K^0},&\ \ \
\bar A_{00}=\vev{\pi^0\pi^0|H|\bar K^0},\cr
A_{+-}=\vev{\pi^+\pi^-|H|K^0},&\ \ \
\bar A_{+-}=\vev{\pi^+\pi^-|H|\bar K^0},\cr
\lambda_{00}=(q/p)(\bar A_{00}/A_{00}),&\ \ \
\lambda_{+-}=(q/p)(\bar A_{+-}/A_{+-}).\cr}}
Then
\eqn\etalam{\eta_{00}={1-\l_{00}\over1+\l_{00}},\ \ \
\eta_{+-}={1-\l_{+-}\over1+\l_{+-}}.}
These quantities get contributions from all three types
of CP violation. It is interesting then to understand
the relative magnitude of each effect and the possibility
of separating them. For this purpose, it is convenient
to discuss $\e$ and $\ep$ instead of $\eta_{00}$ and $\eta_{+-}$.

The $\e$ parameter is defined by
\eqn\defeps{\e\equiv{1\over3}(\eta_{00}+2\eta_{+-})=
{1-\l_0\over1+\l_0},}
where $\l_0$ corresponds to the decay into final $(\pi\pi)_{I=0}$
state, and the second equation holds to first order in $A_2/A_0$.
As, by definition, only one strong channel contributes to $\l_0$,
there is no contribution to \defeps\ from CP violation in decay.
A careful analysis shows that $\Re\e$ is related to CP violation
in mixing ($|q/p|\neq1$) while $\Im\e$ is related to CP violation in
the interference of mixing and decay ($\arg[(q/p)(\bar A_0/A_0)]\neq0$)
\NIR. The two effects are comparable in magnitude.

The $\ep$ parameter is defined by
\eqn\defepsp{\ep\equiv{1\over3}(\eta_{+-}-\eta_{00})\approx{1\over6}
(\l_{00}-\l_{+-}).}
The effect of $|q/p|\neq1$ is negligible, so that to a good approximation
there is no contribution to \defepsp\ from CP violation in mixing.
A careful analysis shows that $\Re\ep$ is related to CP violation in
decay while $\Im\ep$ is related to CP violation in the interference of
mixing and decay \NIR. The two effects are comparable in magnitude.

\subsec{Supersymmetry: Universality and Alignment}

Supersymmetric extensions of the Standard Model predict large new
contributions to FCNC processes. Squarks and gluinos contribute to
$\Delta m_K$ and to $\e$ through box diagrams. A possible suppression
due to large quark and gluino masses is easily compensated for by
three enhancement factors:
\item{(i)} matrix elements of new four-quark operators are enhanced
due to their different Lorentz structure;
\item{(ii)} the weak coupling of the Standard Model diagrams is
replaced by the strong coupling;
\item{(iii)} the GIM mechanism does not operate for generic
squark masses.

The resulting contributions are so large, even for squark masses as
heavy as 1 TeV, that $\Delta m_K$ and $\e$ severely constrain the
form of squark mass matrices
\nref\GaMa{F. Gabbiani and A. Masiero, Nucl. Phys. B322 (1989) 235.}%
\nref\HKT{J.S. Hagelin, S. Kelley and T. Tanaka,
 Nucl. Phys. B415 (1994) 293.}%
\nref\GMS{E. Gabrielli, A. Masiero and L. Silvestrini,
 Phys. Lett. B374 (1996) 80.}%
\nref\GGMS{F. Gabbiani, E. Gabrielli, A. Masiero and L. Silvestrini,
 hep-ph/9604387.}%
\refs{\GaMa-\GGMS}. A convenient way to present these
constraints is the following. Define $K_L^d$ to be the mixing matrix
in the coupling of gluinos to left-handed down quarks and `left-handed'
down squarks and similarly $K_R^d$
(for simplicity, we neglect here L-R mixing among squarks).
Define $\tilde m^2$ to be the average squark mass.
Then, $\Delta m_K$ and $\e$ constrain the following quantities:
\eqn\defdel{(\delta^d_{MM})_{12}\approx(K^d_M)_{11}(K^d_M)_{12}^*\
{\tilde m^2_{\tilde d_{M2}}-\tilde m^2_{\tilde d_{M1}}\over\tilde m^2},
\ \ \ M=L,R.}
With $m_{\tilde q}=m_{\tilde g}=500\ GeV$, ref. \GGMS\ quotes
\eqn\Kconsusy{\eqalign{
\Delta m_K\Longrightarrow&\
\sqrt{\left|\Re(\delta^d_{LL})_{12}^2\right|}\lsim4\times10^{-2},\ \ \
\sqrt{\left|\Re(\delta^d_{LL})_{12}
(\delta^d_{RR})_{12}\right|}\lsim3\times10^{-3};\cr
\e_K\Longrightarrow&\
\sqrt{\left|\Im(\delta^d_{LL})_{12}^2\right|}\lsim3\times10^{-3},\ \ \
\sqrt{\left|\Im(\delta^d_{LL})_{12}
(\delta^d_{RR})_{12}\right|}\lsim2\times10^{-4}.\cr}}

The natural expectation in a generic supersymmetric model
is that mixing angles, mass splittings and phases
are of $\O(1)$, namely $\Re(\delta^d_{MM})_{12}=\O(1)$ and
$\Im(\delta^d_{MM})_{12}=\O(1)$, which would violate \Kconsusy\
by some four orders of magnitude. Two ways of achieving
$(\delta^q_{MM})_{ij}\ll1$ (for $i\neq j$) have been suggested:
\item{1.} {\bf Universality}
\nref\DiGe{S. Dimopoulos and H. Georgi, Nucl. Phys. B193 (1981) 150.}%
\nref\Saka{N. Sakai, Z. Phys. C11 (1981) 153.}%
\refs{\DiGe-\Saka}: The supersymmetry breaking scalar
masses are universal among generations, so that the mass matrices
$\tilde M^2_{d_L}$, $\tilde M^2_{d_R}$ are proportional to 1 and
thus diagonal in any basis. This is achieved in models where the
supersymmetry breaking is communicated by supergravity
\nref\CAN{A.H. Chamseddine, R. Arnowitt and P. Nath,
 Phys. Rev. Lett. 49 (1982) 970; Nucl. Phys. B227 (1983) 1219.}%
\nref\BFS{R. Barbieri, S. Ferrara and C.A. Savoy,
 Phys. Lett. B119 (1982) 343.}%
\nref\HLW{L. Hall, J. Lykken and S. Weinberg,
 Phys. Rev. D27 (1983) 235.}%
\refs{\CAN-\HLW}; in models where supersymmetry is broken at a low scale
and communicated through the Standard Model gauge interactions
\nref\DNS{M. Dine, A. Nelson and Y. Shirman,
 Phys. Rev. D51 (1995) 1362.}%
\nref\DNNS{M. Dine, A. Nelson, Y. Nir and Y. Shirman,
 Phys. Rev. D53 (1996) 2658.}%
\refs{\DNS-\DNNS}; no-scale supergravity and other models
\nref\EKN{J. Ellis, C. Kounnas and D.V. Nanopoulos, Nucl. Phys. B247
 (1984) 373.}%
\nref\LaRo{M. Lanzagorta and G.G. Ross, Phys. Lett. B349 (1995) 319;
 Phys. Lett. B364 (1995) 163.}%
\refs{\EKN-\LaRo}; and (for the first two
generations) in models of non-Abelian horizontal symmetries
\nref\DKL{M. Dine, A. Kagan and R.G. Leigh, Phys. Rev. D48 (1993) 4269.}%
\nref\PoTo{A. Pomarol and D. Tommasini, Nucl. Phys. B466 (1996) 3.}%
\nref\HaMu{L.J. Hall and H. Murayama, Phys. Rev. Lett. 75 (1995) 3985.}%
\refs{\DKL-\HaMu,\BDH}.
\item{2.} {\bf Alignment}
\ref\QSA{Y. Nir and N. Seiberg, Phys. Lett. B309 (1993) 337.}:
The squark mass matrices have a structure, but they have a reason to be
diagonal in the basis set by the quark mass matrix. This is achieved in
models of Abelian horizontal symmetries
\nref\seq{M. Leurer, Y. Nir and N. Seiberg,
 Nucl. Phys. B420 (1994) 468.}%
\refs{\QSA,\seq}\ or dynamically
\ref\DGT{S. Dimopoulos, G.F. Giudice and N. Tetradis,
 Phys. Lett. B357 (1995) 573.}.

Ref. \ref\Pesk{M. Peskin, hep-ph/9604339.}\
describes a systematic experimental program to determine
the mechanism of supersymmetry breaking by direct measurements
in $pp$ and $e^+e^-$ colliders. Here, we wish to show that
FCNC and CP violating processes provide complementary means
of achieving these goals.

The suppression of FCNC and of CP violation is very different
between the two frameworks. If universality holds at the Planck scale,
then at the electroweak scale
\eqn\deluniv{
(K^d_L)_{22}(K^d_L)_{12}^*=V_{cs}V_{cd}^*,\ \ \
{\tilde m^2_{\tilde d_{L2}}-\tilde m^2_{\tilde d_{L1}}\over\tilde m^2}
=\O\left({\ln{m_P\over m_Z}\over16\pi^2}{m_c^2\over m_W^2}\right)
\ \ \Longrightarrow\ \ (\delta^d_{LL})_{12}\sim10^{-5},}
safely below the bounds. (In the $\tilde d_R$ sector, the splittings are
negligible.) On the other hand, in models of alignment,
\eqn\delalig{(K^d_M)_{22}(K^d_M)_{12}^*\sim\sin\theta_c,\ \ \
{\tilde m^2_{\tilde d_{M2}}-\tilde m^2_{\tilde d_{M1}}\over\tilde m^2}
=\O(1)
\ \ \Longrightarrow\ \ (\delta^d_{LL})_{12}\sim10^{-1},}
which is too large. (By ``$\sim$" we mean an order of magnitude estimate
and a possible phase of $\O(1)$.) However, there exist a sub-class of
such models where holomorphy plays an important role and induces
approximate zeros in the down quark mass matrix. As a result, $M^d$ is
very close to being diagonal and the Cabibbo mixing comes from the up
sector. In specific examples in ref. \seq,
\eqn\align{(\delta^d_{LL})_{12}\sim
(K^d_L)_{22}(K^d_L)_{12}^*\sim10^{-4},}
consistent with the constraints from $\Delta m_K$ and
(even with phases of $\O(1)$) from $\e$.

The information from $K$ physics is now built into the various
supersymmetric models, by incorporating either universality or alignment
improved by holomorphy (or a combination of the two mechanisms
\ref\DPS{E. Dudas, S. Pokorski and C.A. Savoy,
 Phys. Lett. B369 (1996) 255.}).
Below we show how measurements of FCNC and/or CP violation in $D$ and
$B$ decays may distinguish between these two possibilities.

\newsec{The $D$ System}

Neither mixing nor CP violation in the $D$ system have been observed.
The Standard Model predicts mixing well below the
experimental bound and negligible CP violation. Therefore, if
mixing is observed in the near future, it will be a clear
signal of New Physics. Below, we explain how $\Delta m_D$ can
potentially play a decisive role in distinguishing between
universality and alignment in supersymmetric theories.
But first we analyze the effects of CP violation on the search
for mixing in the neutral $D$ system.

\subsec{CP Violation in Neutral $D$ decays}
The best bounds on $D-\bar D$ mixing come from measurements
of $D^0\rightarrow K^+\pi^-$
\ref\Anjo{J.C. Anjos {\it et al.}, Phys. Rev. Lett. 60 (1988) 1239.}.
However, these bounds are still
orders of magnitude above the Standard Model prediction for the
mixing. If the value of $\Delta m_D$ is anywhere close to
present bounds, it should be dominated by New Physics.
Then, new CP violating phases may play an important role in
$D-\bar D$ mixing. In this section, we investigate the
consequences of CP violation from New Physics in neutral $D$ mixing
\ref\BSN{G. Blaylock, A. Seiden and Y. Nir,
 Phys. Lett. B355 (1995) 555.}.

As in section 2, we define
\eqn\pqD{|D_{1,2}\rangle=p|D^0\rangle\pm q|\bar D^0\rangle,}
\eqn\AandB{\eqalign{
A\equiv\vev{K^+\pi^-|H|D^0},\ \ \
B\equiv\vev{K^+\pi^-|H|\bar D^0},\cr
\bar A\equiv\vev{K^-\pi^+|H|\bar D^0},\ \ \
\bar B\equiv\vev{K^-\pi^+|H|D^0},\cr}}
\eqn\lD{\lambda={p\over q}{A\over B},\ \ \
\bar\lambda={q\over p}{\bar A\over\bar B}.}

The following approximations can be safely made:
\item{(i)} $\Delta M\ll\Gamma$, $\Delta\Gamma\ll\Gamma$, $|\lambda|\ll1$
(all experimentally confirmed).
\item{(ii)} $\Delta\Gamma\ll\Delta M$
(which is very likely if $\Delta M$ is close to the bound).

We further make the following very reasonable assumptions:
\item{(iii)} CP violation in decay is negligible,
$|A/\bar A|=|B/\bar B|=1$.
\item{(iv)} CP violation in mixing is negligible, $|p/q|=1$.

The two assumptions together imply also $|\lambda|=|\bar\lambda|$.

The consequence of $(i)-(iv)$ is the following
form for the (time dependent) ratio between the
doubly Cabibbo suppressed (DCS) and
Cabibbo-allowed decay rates ($D^0(t)$ [$\bar D^0(t)$] is the
time-evolved initially pure $D^0$ [$\bar D^0$] state):
\eqn\Drate{\eqalign{
{\Gamma[D^0(t)\rightarrow K^+\pi^-]\over
\Gamma[D^0(t)\rightarrow K^-\pi^+]}=&\
|\lambda|^2+{\Delta M^2\over4}t^2+{\rm Im}(\lambda)\ t,\cr
{\Gamma[\bar D^0(t)\rightarrow K^-\pi^+]\over
\Gamma[\bar D^0(t)\rightarrow K^+\pi^-]}=&\
|\lambda|^2+{\Delta M^2\over4}t^2+{\rm Im}(\bar\lambda)\ t.\cr}}
This form is valid for time $t$ not much larger than ${1\over\Gamma}$.
The time independent term is the DCS decay contribution;
the term quadratic in time is the pure mixing contribution;
and the term linear in time results from the interference
between the DCS decay and the mixing amplitudes.
Note that both the const($t$) and the $t^2$ terms are equal
in the $D^0$ and $\bar D^0$ decays. However, if $CP$ violation
in the interference of mixing and decay is significant,
${\rm Im}(\lambda)\neq{\rm Im}(\bar\lambda)$
and the linear term is different for $D^0$ and $\bar D^0$.

The experimental strategy should then be as follows:
(a) Measure $D^0$ and $\bar D^0$ decays separately.
(b) Fit each of the ratios to constant plus linear plus quadratic
time dependence.
(c) Combine the results for $|\lambda|^2$ and $\Delta M^2$.
(d) Compare Im($\lambda$) to Im($\bar\lambda$).

The comparison of the linear term should be very informative
about the interplay between strong and weak phases in these decays.
There are four possible results:
\item{1.} ${\rm Im}(\lambda)={\rm Im}(\bar\lambda)=0$:
Both strong phases and weak phases play no role in these processes.
\item{2.} ${\rm Im}(\lambda)={\rm Im}(\bar\lambda)\neq0$:
Weak phases play no role in these processes. There is a different
strong phase shift in $D^0\rightarrow K^+\pi^-$ and
$D^0\rightarrow K^-\pi^+$.
\item{3.} ${\rm Im}(\lambda)=-{\rm Im}(\bar\lambda)$:
Strong phases play no role in these processes. CP violating phases
affect the mixing amplitude.
\item{4.} $|{\rm Im}(\lambda)|\neq|{\rm Im}(\bar\lambda)|$:
Both strong phases and weak phases play a role in these processes.

In all these cases, the magnitude of the strong and the weak phases
can be determined from the values of $|\lambda|$, Im($\lambda)$
and Im($\bar\lambda$).

Finding either quadratic or linear time dependence would be
a signal for mixing in the neutral $D$ system. However,
a non-vanishing linear term does not by itself signal CP violation
in mixing, only if it is different in $D^0$ and $\bar D^0$.
The linear term could be a problem for experiments: if the phase
is such that the interference is destructive, it could partially
cancel the quadratic term in the relevant range of time, thus
weakening the experimental sensitivity to mixing \BSN. On the other
hand, if the mixing amplitude is smaller than the DCS one,
the interference term may signal mixing even if the pure mixing
contribution is below the experimental sensitivity
\nref\Liu{T. Liu, hep-ph/9408330.}%
\nref\Wol{L. Wolfenstein, Phys. Rev. Lett. 75 (1995) 2460.}%
\refs{\Liu-\Wol}.

\subsec{Supersymmetry: Universality and Alignment}
The constraints from $\Delta m_D$ analogous to  \Kconsusy\ are \GGMS:
\eqn\Dconsusy{\Delta m_D\Longrightarrow\
\sqrt{\left|\Re(\delta^u_{LL})_{12}^2\right|}\lsim1\times10^{-1},\ \ \
\sqrt{\left|\Re(\delta^u_{LL})_{12}
(\delta^u_{RR})_{12}\right|}\lsim2\times10^{-2}.}
In models of universality,
\eqn\delunivD{
(K^u_L)_{22}(K^u_L)_{12}^*=\O\left(
{\ln{m_P\over m_Z}\over16\pi^2}\right)V_{us}V_{cs}^*,\ \ \
{\tilde m^2_{\tilde u_{L2}}-\tilde m^2_{\tilde u_{L1}}\over\tilde m^2}
=\O\left({m_c^2\over m_W^2}\right)
\ \ \Longrightarrow\ \ (\delta^u_{LL})_{12}\sim10^{-5},}
safely below the bounds. On the other hand, in models with alignment,
if -- as required by the $K$ system and achievable with holomorphy --
$(K^d_L)_{12}\ll\sin\theta_c$, then necessarily  \QSA\
\eqn\delaligD{(K^u_L)_{22}(K^u_L)_{12}^*\sim\sin\theta_c
\ \ \Longrightarrow\ \ (\delta^u_{LL})_{12}\sim10^{-1},}
(we take the mass splitting to be of $\O(1)$).
{\it Models of quark-squark alignment predict that $\Delta m_D$ is
close to the experimental bound.} Furthermore, the supersymmetric
contribution to the mixing could carry a new, large CP violating
phase. Such a phase has interesting implications for the search of
$D-\bar D$ mixing, as described in the previous subsection.

\newsec{The $B$ System}

\nref\NiQurev{For a review, see Y. Nir and H.R. Quinn,
 Ann. Rev. Nucl. Part. Sci. 42 (1992) 211.}%
\subsec{Beyond the Standard Model - General \NiQurev}

CP asymmetries in $B$ decays are a sensitive probe of new physics
in the quark sector, because they are likely to differ from the
Standard Model predictions if there are sources of CP violation
beyond the CKM phase of the Standard Model. New Physics can contribute
in two ways:
\item{(i)} If there are significant contributions to $B-\bar B$
mixing (or $B_s-\bar B_s$ mixing) beyond the box diagram
with intermediate top quarks; or
\item{(ii)} If the unitarity of the three-generation CKM matrix
does not hold, namely if there are additional quarks.
\par Actually, there is a third way in which the Standard Model
predictions may be modified even if there are no new sources
of CP violation:
\item{(iii)} The constraints on the CKM parameters change if there are
significant new contributions to $B-\bar B$ mixing and to $\epsilon_K$
(see e.g.
\ref\GrNi{Y. Grossman and Y. Nir, Phys. Lett. B313 (1993) 126.}).

On the other hand, the following ingredients of the analysis of
CP asymmetries in neutral $B$ decays are likely to hold in most
extensions of the Standard Model:
\item{(iv)} $\Gamma_{12}\ll M_{12}$. This is not only theoretically very
likely but also supported by experimental evidence: $\Delta M/\Gamma
\sim0.7\ (\gsim6)$ for $B_d\ (B_s)$, while branching ratios into states
that contribute to $\Gamma_{12}$ are $\leq10^{-3}\ (0.1)$.
\item{(v)} The relevant decay processes (for tree decays) are dominated
by Standard Model diagrams. It is unlikely that new physics, which
typically takes place at a high energy scale, would compete with weak
tree decays. (On the other hand, for penguin dominated decays, there
could be significant contributions from new physics.)

Within the Standard Model, both $B$ decays and $B-\bar B$ mixing are
determined by combinations of CKM elements. The asymmetries then
measure the relative phase between these combinations. Unitarity of
the CKM matrix directly relates these phases (and consequently
the measured asymmetries) to angles of the unitarity triangles.
In models with new physics, unitarity of the three-generation
charged-current mixing matrix may be lost and consequently the
relation between the CKM phases and angles of the unitarity triangle
violated. But this is not the main reason that the predictions for the
asymmetries are modified. The reason is rather that if $B-\bar B$
mixing has significant contributions from new physics, the asymmetries
measure different quantities: the relative phases between the CKM
elements that determine $B$ decays and the elements of mixing matrices
in sectors of new physics (squarks, multi-scalar, etc) that contribute
to $B-\bar B$ mixing.

Thus, when studying CP asymmetries in models of new physics, we look
for violation of the unitarity constraints and, more importantly,
for contributions to $B-\bar B$ mixing that are different
in phase and not much smaller in magnitude than the Standard Model
contribution. In Supersymmetry, the aspect of new CP violating phases
in $B-\bar B$ mixing is markedly different in the cases of universality
and alignment. We explain this point in the next subsection.

\subsec{Supersymmetry: Universality and Alignment}

The constraints from $\Delta m_B$ analogous to  \Kconsusy\ are \GGMS:
\eqn\Dconsusy{
\Delta m_B\Longrightarrow\
\sqrt{\left|\Re(\delta^d_{LL})_{13}^2\right|}\lsim1\times10^{-1},\ \ \
\sqrt{\left|\Re(\delta^d_{LL})_{13}
(\delta^d_{RR})_{13}\right|}\lsim2\times10^{-2}.}
In models of universality,
\eqn\delunivD{(K^d_L)_{33}(K^d_L)_{13}^*=V_{td}V_{tb}^*,\ \ \
{\tilde m^2_{\tilde d_{L3}}-\tilde m^2_{\tilde d_{L3}}\over\tilde m^2}
=\O\left({\ln{m_P\over m_Z}\over16\pi^2}{m_t^2\over m_Z^2}\right)
\ \ \Longrightarrow\ \ (\delta^d_{LL})_{13}\sim10^{-3}.}
In models with alignment,
\eqn\delaligD{(K^d_L)_{33}(K^d_L)_{13}^*\sim V_{td}V_{tb}^*
\ \ \Longrightarrow\ \ (\delta^d_{LL})_{13}\sim10^{-2}.}
A supersymmetric contribution
to $B-\bar B$ mixing of $\O(0.1)$ is possible. The crucial difference
between universality and alignment does not lie, however, in the
magnitude of the contributions: these are too small to be clearly
signalled in $\Delta m_B$ because of the hadronic uncertainties
(most noticeably in $f_B$). The important difference lies in the fact
the the supersymmetric contribution in the models of universality
carries the same phase as the Standard Model box diagram, while
in models of alignment the phase is unknown. Therefore,
models of universality predict no effect on CP asymmetries
in $B$ decays, while models of alignment allow reasonably
large deviations from the Standard Model.

\newsec{Conclusions}
FCNC and CP violation in the $K$ system have played an extremely
important role in shaping the way we think about supersymmetry.
In particular, to solve the supersymmetric flavor problems,
all models incorporate either universality or alignment.
Future measurements of mixing and CP violation should allow us
to distinguish between the two possibilities:
\item{a.} Alignment predicts that $D-\bar D$ mixing is close
to the present experimental bound. Universality predicts that it
is well below the bound.
\item{b.} Alignment allows large CP violation in $D-\bar D$
mixing. Universality predicts that it is negligible.
\item{c.} Alignment allows shifts in CP asymmetries in neutral $B$
decays into final CP eigenstates (compared to the Standard Model
contributions) of order 0.2. Universality does not modify the
Standard Model values.

The combination of these measurements might then exclude or strongly
support either of these supersymmetric frameworks.

\vskip 1cm
{\bf Acknowledgments:}
I thank Francesca Borzumati and Yuval Grossman for their help
in preparing this talk.
Y.N. is supported in part by the United States -- Israel Binational
Science Foundation (BSF), by the Israel Commission for Basic Research,
and by the Minerva Foundation (Munich).

\listrefs
\end